# Soliton-like magnetic domain wall motion induced by the interfacial Dzyaloshinskii-Moriya interaction

Yoko Yoshimura[1], Kab-Jin Kim[1], Takuya Taniguchi[1], Takayuki Tono[1], Kohei Ueda[1], Ryo Hiramatsu[1], Takahiro Moriyama[1], Keisuke Yamada[2], Yoshinobu Nakatani[2], and Teruo Ono[1]

[1]*Institute for Chemical Research, Kyoto University, Gokasho, Uji, Kyoto, 611-0011, Japan*

[2]*University of Electro-communications, Chofu, Tokyo, 182-8585, Japan*

Topological defects such as magnetic solitons, vortices, Bloch lines, and skyrmions have started to play an important role in modern magnetism because of their extraordinary stability, which can be exploited in the production of memory devices. Recently, a novel type of antisymmetric exchange interaction, namely the Dzyaloshinskii-Moriya interaction (DMI), has been uncovered and found to influence the formation of topological defects. Exploring how the DMI affects the dynamics of topological defects is therefore an important task. Here we investigate the dynamic domain wall (DW) under a strong DMI and find that the DMI induces an annihilation of topological vertical Bloch lines (VBLs) by lifting the four-fold degeneracy of the VBL. As a result, velocity reduction originating from the Walker breakdown is completely suppressed, leading to a soliton-like constant velocity of the DW. Furthermore, the strength of the DMI, which is the key factor for soliton-like DW motion, can be quantified without any side effects possibly arising from current-induced torques or extrinsic pinnings in magnetic films. Our results therefore shed light on the physics of dynamic topological defects, which paves the way for future work in topology-based memory applications.

The magnetic domain wall (DW), also known as the magnetic soliton, has received significant attention because of the academic interest it inspires, as well as its potential applications in data storage and logic devices[1–12]. The dynamics of the DW consists of unique, nonlinear behaviour in response to an external magnetic field. In one-dimensional wires, where the degree of freedom along the wire's transverse direction is suppressed, the DW velocity increases linearly with the external magnetic field up to a threshold, beyond which it abruptly decreases. The abrupt reduction of DW velocity is due to the onset of precessional DW motion, which causes a periodic change in the DW's helicity [Fig. 1a and Supplementary Fig. S5]. This is a well-known phenomenon in field-driven DW dynamics, and is referred to as the Walker breakdown (WB)[1,8]. An actual DW, however, may have two-dimensional configurations, where the coherent precessional motion of the DW above the threshold field (hereafter the Walker field) should be replaced by the nucleation and a propagation of vertical Bloch lines (VBLs) [Fig. 1b][1].

The VBL, a magnetic curling structure that divides the DW, is considered to be a topological defect as its stability can be determined based on a topological argument. The VBL has four-fold degeneracy that depends on the magnetic charge ($Q = \pm 1$) and chirality ($C = \pm 1/2$). Here, the topological charge corresponds to the sign of magnetic charge at the centre of the VBL – that is to say, $Q = +1$ for head-to-head spin alignment and $Q = -1$ for tail-to-tail spin alignment. Positive (negative) chirality is defined as clockwise (counter-clockwise) rotation of the spin. The half-integer feature of the chirality indicates the half-cycle of the spin rotation (so called π-VBL). We define the four degenerate states of VBL as shown in Fig. 1c. The four degenerate states all generally share the same energy level, and thus appear with equal probability[3]. Above the Walker field, VBLs are continuously nucleated and are driven out of the sample through the edge [Fig. 1b], resulting in a reduced average DW velocity, which is analogous to the WB phenomenon in a one-dimensional wire.

The velocity breakdown due to the WB generally limits the functional performance in DW motion-based devices, and thus efforts have been made toward avoiding it. To suppress the WB, several proposals have been suggested based on micromagnetics simulations[9–12]. In the proposed methods, however, a particular geometry is required to suppress the WB. By way of contrast, here we report on our experimental discovery that the WB can be completely suppressed under the strong Dzyaloshinskii-Moriya interaction (DMI), which inherently exists at the interface in ferromagnetic/nonmagnetic bi-layer structures[13–17]. Such a suppression of the WB is observed in two-dimensional wires because the evolution of the topological VBL plays a key role on the suppression of the WB. The DMI is found to lift the degeneracy of VBLs and split their energy; this energy splitting yields a remarkable difference in the velocity of VBLs, leading to a unidirectional collision of pairs of VBLs. In turn, this unidirectional collision of VBLs induces a locking of the average azimuthal angle of the DW even beyond the Walker field, leading to a soliton-like constant velocity far above the Walker field. The annihilation process of two topological VBLs releases a large amount of energy via spin wave emission. Such an energy relaxation can balance the Zeeman energy relaxation rate in a high-field regime, thereby accounting for the suppression of the WB in terms of energy. The strength of the DMI, which is a key factor in understanding the soliton-like DW motion, is also quantified experimentally. These results not only provide a novel way to suppress the WB phenomenon, but they also shed light on the underlying physics of two-dimensional soliton-like DW dynamics.

For this study, two types of perpendicularly magnetized Co/Ni films – i.e. symmetric and asymmetric layers – were prepared. For the symmetric Co/Ni sample, we used Si/Ta (4 nm)/Pt (2 nm)/Co (0.3 nm)/Ni (0.6 nm)/Co (0.3 nm)/Pt (2 nm)/Ta (4 nm), for which the interfacial DMI should be cancelled out owing to the structural inversion symmetry. On the other hand, Si/Ta (4 nm)/Pt (2 nm)/Co (0.3 nm)/Ni (0.6 nm)/Co (0.3 nm)/MgO (1 nm)/Pt (2 nm)/Ta (4 nm) was used for the asymmetric Co/Ni sample, with MgO being inserted between the upper Co and Pt layers in order to break the structural inversion symmetry; accordingly, the interfacial DMI should exist in the asymmetric Co/Ni layer[17–19]. 1 μm wide and 50 μm long wires are fabricated by electron beam lithography and Ar ion milling, as shown in Fig. 2a. Two Au (100 nm)/Ta (5 nm) contact lines, labelled A and B, were attached to the wire in order to flow direct current into the wire, which is necessary for measuring the Hall voltage. A 100 nm wide Hall bar structure was designed to detect DW motion through the anomalous Hall voltage $V_H$. To investigate the DW velocity in a flow regime, we developed a time-of-flight measurement of DW propagation. The details of the measurement technique and the measurement procedure are described in the Supplementary Information.

Figures 2b and 2c show the DW velocity as a function of |$H$| for symmetric and asymmetric Co/Ni wires, respectively. A clear threshold field originating from the pinning of wires is observed, implying that the DW velocities measured over the threshold field belong to the flow regime[20]. We observed a distinct difference in the DW velocity between symmetric and asymmetric Co/Ni wires. The DW velocity of the asymmetric wire is much larger than that of the symmetric wire, and it keeps almost constant above 150 mT. On the other hand, the DW velocity in the symmetric wire is comparatively smaller and slightly increases with the magnetic field.

To elucidate the underlying mechanism, we first check the Walker field in our sample based on the one-dimensional DW model including the interfacial DMI parameter $D$. Here, the interfacial DMI is assumed to originate from the lack of inversion symmetry at the interface, yielding a chiral interaction between two atomic spins $S_1$ and $S_2$, which can be expressed as the Hamiltonian: $H_{DM} = -\boldsymbol{D} \cdot (\boldsymbol{S}_1 \times \boldsymbol{S}_2)$[13,14]. We note that the Walker field in two-dimensional DW motion is the same as that in one-dimensional DW motion[21]. The calculated DW velocities are plotted as solid lines in Figs. 2b and 2c for various values of $D$. It is clear that the WB field is far below the pinning field of our samples, implying that the WB phenomenon is obscured by the DW creep motion that results from pinnings[20,22,23]. This implies that the DW velocities observed in Figs. 2b and 2c belong to the precessional regime.

The dynamics of DW precession depends on the dimensions of the wire. In narrow wires, a coherent DW precession occurs by periodically overcoming the DW anisotropy energy [Fig. 1a]. However, if the wire width is larger than the width of the VBL ($\Lambda$), we intuitively expect that it is sufficient to nucleate VBLs since the energy consumption for nucleating VBLs (mainly due to the increase of exchange energy) becomes smaller than that for the coherent precession [Fig. 1b][1]. Here, $\Lambda$ is determined by the exchange energy ($A$) and the DW anisotropy energy ($K_d$): $\Lambda = \pi\sqrt{A/K_d}$.

Micromagnetics simulations were performed to understand the experimental results. Two-dimensional wires including a pinning distribution were used for the simulation. The details of the simulation are described in the Methods section. Figures 2d and 2e show the simulation results of DW velocity for various values of *D* by using the material parameters for symmetric and asymmetric Co/Ni wires, respectively. The simulations reproduce the experimental results quite well by assuming that $D = 0.14$ mJ/m$^2$ and $D = 0.6$ mJ/m$^2$ for symmetric and asymmetric wires, respectively. We performed further simulations in an ideal wire to escape the complexities that result from the pinnings. Figure 2f shows the DW velocity as a function of the out-of-plane magnetic field for various values of *D*. In the absence of the DMI (i.e. $D = 0$ mJ/m$^2$), the DW velocity follows the conventional Walker rigid body model, which predicts a velocity breakdown above the Walker field [see the inset of Fig. 2f][8]. Interestingly, however, as the DMI increases, such a velocity breakdown is gradually suppressed and the DW velocity remains constant for a wide range of magnetic fields, even above the WB field.

The time evolution of the magnetization distribution of the DW sheds light on the effect of the DMI on the DW motion. Figures 3a and 3b show snapshots of a moving DW above the Walker field ($|H| = 15$ mT for $D = 0$ mJ/m$^2$ and $|H| = 150$ mT for $D = 1.0$ mJ/m$^2$) [see also Supplementary Movie 1 and 2]. It is clearly shown that the VBLs appear inside the DW, indicating the two-dimensional nature of the wire. Noticeably, the dynamics of the VBLs strongly depends on the DMI. For $D = 0$ mJ/m$^2$, the VBLs nucleated inside the DW propagate to the edge [see Fig. 3c]. On the other hand, for the case of $D = 1.0$ mJ/m$^2$, the VBLs nucleated inside the DW annihilate immediately by emitting spin waves [see Fig. 3d]. This intriguing feature can be understood based on the topological features of VBLs.

Figure 4a shows the DW anisotropy energy profile as a function of $\varphi$. Here, $\varphi$ is the azimuthal angle of the DW defined in the *x-y* plane, as denoted in Fig. 1b. The DW anisotropy energy ($\sim\cos^2\varphi$) has two equivalent minima at $\varphi = \pm\pi/2$ for one cycle of spin rotation ($-\pi < \varphi < \pi$); that is to say, the spin aligned along the $+y$ direction is energetically equivalent to that aligned along the $-y$ direction. Since the spins can rotate in opposite directions with equal probability, the VBL has four degenerate states with different topological charges ($Q = \pm 1$) and chiralities ($C = \pm 1/2$)[1,3,24] as shown in Fig. 1c.

The evolution of the VBL (nucleation, propagation, and annihilation processes) is governed by topological constraints. During the nucleation or annihilation process, the total topological charge should remain constant, i.e. $\sum_i Q_i^{before} = \sum_i Q_i^{after}$ [25,26]. The total chirality ($\sum_i C_i$) is referred to as the 'winding number $w$', which counts how many times the spin is wrapped around the circle through the DW. The winding number is also known as a topological number because configurations with different winding numbers cannot be continuously deformed into each other[27]. Therefore, the winding number also should remain constant during the nucleation and annihilation processes. Note that the topological argument does not hold at the edge because the VBLs may simply be driven out of the sample through the edges or injected from the edge [see Supplementary Information].

Figure 4b shows an example of the evolution of VBLs in the absence of the DMI. Two VBLs with opposite charge and chirality are nucleated simultaneously in order to satisfy the topological constraints. Because they have equal energy, they propagate along the DW with equal velocity. Note that the propagation direction is determined by the sign of the chirality. Successive nucleation and propagation of VBLs induces a local spin precession (marked by purple arrows), which is analogous to the precessional motion of a DW in a one-dimensional wire.

In the presence of the DMI, however, the topological characteristics of the DW are quite different. The DMI energy, unlike the DW anisotropy energy, has cos $\varphi$ symmetry, which has only one single energy minimum for $-\pi < \varphi < \pi$[21]. A schematic illustration of the energy profile is shown in Fig. 4a. When the DMI energy and DW anisotropy energy coexist, the DMI energy distorts the periodic nature of the DW anisotropy energy and lifts the degeneracy of the VBL, splitting the energy levels into two sub-levels, as shown in Fig. 4c.

Such an energy splitting of the VBL influences the dynamics of DW considerably. As a result of the energy splitting, the width of the VBL in two energy levels becomes different. Here, the width of the VBL is defied by $\Lambda = \pi\sqrt{A/K_d^{eff}}$, where $K_d^{eff}$ is an effective DW anisotropy energy including the DMI energy. Considering that the velocity of the VBL is proportional to its width[24], the VBLs in the ground state become much faster than those in the excited state. Such a velocity difference induces a unidirectional collision of two VBLs[28]. The detailed procedure of unidirectional collision is shown in Fig. 4d. Two VBLs with opposite charge and the same chirality move to the same directions as those in Fig. 4b. (VBLs marked by green and red colours in Fig. 4d). However, owing to the DMI-induced energy splitting, the velocities of the two VBLs are quite different, and they finally collide with each other. This is the underlying reason why we did not observe the propagation of VBLs in Fig. 3d.

Here, importantly, the unidirectional collision violates the topological constraint in the winding number since the winding number changes from $w = 1$ to $w = 0$ during collision. Such an annihilation process that changes the winding number is only permitted at the edge [see Supplementary Information]. Thus, the annihilation of the VBL inside the DW is required to release a large amount of energy, which can be achieved via spin wave emissions, as shown in Fig. 3b.

Interestingly, the unidirectional collision of two VBLs suppresses the local spin precession (compare the purple arrows in Fig. 4b and 4d). This results in an important fact: namely, that the continuous annihilation of VBLs induces another constraint on the DW. More specifically, the average azimuthal angle of DW ($\varphi$) should remain constant even beyond the Walker field. Figure 4e shows the time evolution of $\varphi$ above the WB field ($|H|$ = 15 mT for $D$ = 0 mJ/m$^2$ (red line) and $|H|$ = 150 mT for $D$ = 1.0 mJ/m$^2$ (blue line)). Here, $\varphi$ is obtained by averaging the angle of the total magnetic moment inside the DW. Without the DMI, $\varphi$ shows a continuous increase with time, corresponding to the precessional motion of DW. On the other hand, under a strong DMI, the change of $\varphi$ is suppressed and $\varphi$ is fixed around $\pi/2$. In Fig. 4e, we also plot $\varphi$ for $|H|$ = 75 mT (green line) just before the WB field, in which the DW reaches peak velocity without precession [see Supplementary Movie 3]. The results shows that the average angle $\varphi$ is almost the same for $|H|$ = 75 mT and 150 mT. This explains why the DW velocity remains constant at a peak velocity under a strong DMI [Fig. 2f]. In other words, despite the apparently dissimilar appearance of DWs for $|H|$ = 75 mT and for above the WB field, they all exhibit the same average angle $\varphi$, which yields the same velocity. This line of thought is analogous to that regarding the magnetic soliton, in which the DW velocity is simply determined by the azimuthal angle of the DW[27]. Interestingly, however, here we find that soliton-like DW velocity can be maintained even above the WB field in two-dimensional DW configurations. In addition, our results imply that the helical symmetry of the DW is broken as a result of the DMI (see Supplementary Information).

The suppression of the Walker breakdown can be supported by energy considerations in the system. The saturation in the DW velocity above the WB field [Fig. 2f] implies that another energy dissipation channel is opened because the dissipation via Gilbert damping ($\alpha$) cannot follow up the Zeeman energy relaxation rate for such a fast DW motion for a high-field regime[9,29]. We ascribe the additional energy relaxation channel to the local spin wave emission induced by the unidirectional collision of VBLs, because the annihilation of topologically protected defects generally accompanies a large energy relaxation[27]. Such a spin wave emission can balance with the Zeeman energy relaxation rate for fast DW motion[29,30]. We confirm this using micromagnetic simulations with $\alpha = 0$. Even for $\alpha = 0$, the DW is found to move in two-dimensional wires by emitting spin waves [see Supplementary Movie 4]. Note that we do not observe any DW motion in one-dimensional wires for $\alpha = 0$ even under a strong DMI. This highlights the fact that the evolution of the VBL is crucial for the DW motion under the DMI. The constant DW velocity across a wide range of out-of-plane fields can be interpreted to mean that the number of VBLs increases with the field in order to compensate for the increasing Zeeman energy relaxation rate. The increase in the number of VBLs is indeed observed in our simulation.

The strength of the DMI is a key factor in understanding the soliton-like two-dimensional DW motion. We can quantify it by measuring the DW velocity as sweeping an in-plane longitudinal field because the energy profile induced by the DMI is identical to the Zeeman energy caused by an in-plane longitudinal field. Figure 5a shows the simulated DW velocity as a function of the in-plane field with various values of $D$. For $D = 0$ mJ/m$^2$, the DW velocity has a minimum at $H_{\min} = 0$ Oe. The parallel shift of $H_{\min}$ is observed as the DMI increases, indicating that the effect of the DMI can be cancelled out by applying an appropriate in-plane longitudinal field.

Figure 5b shows the plot of $H_{\min}$ versus $D$. We also plot the $H_{\mathrm{DMI}}$ versus $D$ (red line) based on $H_{\mathrm{DMI}} = \dfrac{D}{M_S \Delta}$. Here, $H_{\mathrm{DMI}}$ is the DMI-induced effective magnetic field[21], $M_S$ the saturation magnetization and $\Delta$ the DW width parameter. Exact coincidence between $H_{\min}$ and $H_{\mathrm{DMI}}$ suggests that the DMI can be quantified by achieving minimum DW velocity by sweeping the in-plane field. Note that any ambiguities possibly originating from the current-induced torques or pinnings in the wires can be ruled out in this scheme because of its field-driven dynamic nature in a flow regime. Figures 5c and 5d show experimental results of the in-plane field dependence of the DW velocity obtained from symmetric and asymmetric Co/Ni wires, respectively. For the symmetric Co/Ni wire, the $H_{min}$ is observed at $|H_{min}| = 12.0 \pm 0.4$ mT, which corresponds to a DMI strength of $0.038 \pm 0.001$ mJ/m$^2$. The finite $D$ in the symmetric Co/Ni wire possibly arises from the dissimilar interface between the bottom Pt/Co and the top Co/Pt[31]. For the asymmetric Co/Ni wire, $H_{min}$ is estimated to be $|H_{min}| = 213 \pm 55$ mT by extrapolation, which corresponds to a DMI strength of $0.6 \pm 0.15$ mJ/m$^2$, which is consistent with the simulation results [Fig. 2e].

In conclusion, we demonstrated that the DMI suppresses the WB phenomenon in two-dimensional DW motion, and that it yields soliton-like DW velocity. It was found that the evolution of topological VBLs plays a crucial role in this soliton-like DW motion. The suppression of the Walker breakdown was supported by energy considerations of the system. In addition, we provide a fundamentally new method of measuring the DMI strength without any side effects possibly arising from the current-induced torques or extrinsic pinnings in magnetic films. These results not only shed light on the underlying physics of two-dimensional DW motion involving the DMI, but they also provide a new scheme for quantifying the DMI, which can act as a fundamental building block in emerging magnetic devices.

## Methods

**Film preparation and device fabrication.** Symmetric and asymmetric Co/Ni films were deposited on an undoped Si substrate by DC magnetron sputtering. The structure of the symmetric Co/Ni film was as follows: Si/Ta (4 nm)/Pt (2 nm)/Co (0.3 nm)/Ni (0.6 nm)/Co (0.3 nm)/Pt (2 nm)/Ta (4 nm) . For asymmetric Co/Ni films, we inserted an MgO (1 nm) layer between the upper Pt and Co layers. High-quality Co/Ni films were obtained by using a deposition rate of 0.73 Å/s through adjustments to the Ar sputtering pressure (2.7 mPa) and sputtering power (300 W). Co/Ni wires with a 100 nm wide Hall cross structure were fabricated using electron beam lithography and Ar ion milling. A negative tone electron beam resist (maN-2403) was used for lithography at a fine resolution (~5 nm). For current injection, two electrodes were stacked on each nanowire, as shown in Fig. 2a. To make an Ohmic contact, the surface of the nanowire was cleaned by weak ion milling before electrode deposition.

**Experimental set up.** The configuration of the circuit employed is illustrated in Fig. 2a. A pulse generator (Picosecond 10, 300B) was used to generate a current pulse to create the DW. The current pulse propagates through contact line A and is recorded in the oscilloscope (Textronix 7354). The voltage signal originating from the Hall cross is also recorded in the oscilloscope through the 46 dB differential amplifier. The simultaneous recording of the two signals (i.e. the current pulse for creating the DW and the Hall voltage for detecting the DW) allowed us to measure the DW arrival time, which can be converted into the DW's velocity. A direct current source (Yokogawa 7651, max 30 V, 100 mA) was used to inject direct current to generate the anomalous Hall voltage.

**Micromagnetics simulations.** All micromagnetics simulations were performed using a program developed previously[9,32]. A 500 nm wide wire was used for the simulation. The sample was divided into identical rectangular prisms (cells), in which the magnetization was assumed to be constant. The cross-sectional dimensions of the wire in the simulation were $500 \times 1.2$ nm$^2$. The moving calculation region, always centred on the DW, was limited to a length of 1 μm along the wire. The exchange field was evaluated from the four neighbouring cells. The demagnetizing field was averaged over the entire cell[33], and was evaluated using the fast Fourier transform technique with zero padding in order to improve calculation speed[34].

Furthermore, the anisotropy distribution was taken into account in the simulation to reproduce the DW pinning field. The DMI energy originating from the antisymmetric exchange interaction was included together with the other micromagnetic energies. A boundary condition that took into account the bending of the magnetization at the edges that resulted from the DMI was used[35]. The two-dimensional calculations were performed by dividing the wire into rectangular prisms of $2 \times 2 \times 1.2$ nm$^3$. The time step was 0.25 ps. The DW velocity was determined over the course of 16 simulations for each magnetic field. The experimentally determined material parameters used were as follows: saturation magnetization $M_S = 8.37 \times 10^5$ A/m; perpendicular magnetic anisotropy $K_U = 0.9 \times 10^6$ J/m$^3$ (for symmetric Co/Ni) and $1.31 \times 10^6$ J/m$^3$ (for asymmetric Co/Ni); damping constant[36] $\alpha = 0.15$.

# References

1. Malozemoff, A. P. & Slonczewski, J. C. *Magnetic domain walls in bubble materials* (Academic Press, New York, 1979).

2. Mikeska, H. J. Solitons in a one-dimensional magnet with an easy plane. *J. Phys. C: Solid State Phys.* **11**, L29–32 (1978).

3. Hubert, A. & Schäfer, R. *Magnetic Domains* (Springer, Berlin, 1998).

4. Ono, T. *et al*. Propagation of a magnetic domain wall in a submicrometer magnetic wire. *Science* **284**, 468 (1999).

5. Yamaguchi, A. Real-space observation of current-driven domain wall motion in submicron magnetic Wires. *Phys. Rev. Lett*. **92**, 077205 (2004).

6. Allwood, D. A. *et al*. Magnetic domain-wall logic. *Science* **309**, 1688–1692 (2005).

7. Parkin, S. S. P., Hayashi, M., & Thomas, L. Magnetic domain-wall racetrack memory. *Science* **320**, 190 (2008).

8. Schryer, N. L. & Walker, L. R. The motion of 180° domain walls in uniform dc magnetic fields. *J. Appl. Phys.* **45**, 5406–5421 (1974).

9. Nakatani, Y., Thiaville, A. & Miltat, J. Faster magnetic walls in rough wires. *Nat. Mater.* **2**, 521–523 (2003).

10. Lee, J. Y., Lee, K. S. & Kim, S. K. Remarkable enhancement of domain-wall velocity in magnetic nanostripes. *Appl. Phys. Lett.* **91**, 122513 (2007).

11. Yan, M., Andreas, C., Kákay, A., Garcia-Sánchez, F. & Hertel, R. Fast domain wall dynamics in magnetic nanotubes: Suppression of Walker breakdown and Cherekov-like spin wave emission. *Appl. Phys. Lett*. **99**, 122505 (2011).

12. Burn, D. M. & Atkinson, D. Suppression of Walker breakdown in magnetic domain wall propagation through structural control of spin wave emission. *Appl. Phys. Lett.* **102**, 242414 (2013).

13. Dzyaloshinskii, I. A thermodynamic theory of "weak" ferromagnetism of antiferromagnetics. *J. Phys. Chem. Solids* **4**, 241–255 (1958).

14. Moriya, T. Anisotropic superexchange interaction and weak ferromagnetism. *Phys. Rev.* **120**, 91–98 (1960).

15. Emori, S., Bauer, U., Ahn, S. –M., Martinez, E. & Beach, G. S. D. Current-driven dynamics of chiral ferromagnetic domain walls. *Nat. Mater*. **12**, 611–616 (2013).

16. Ryu, K.-S., Thomas, L., Yang, S.-H. & Parkin, S. Chiral spin torque at magnetic domain walls. *Nat. Nanotechnol.* **8**, 527–533 (2013).

17. Ueda, K. *et al*. Transition in mechanism for current-driven magnetic domain wall dynamics. *Appl. Phys. Express* **7**, 053006 (2014).

18. Kim, K.-J. *et al*. Tradeoff between low-power operation and thermal stability in magnetic domain-wall-motion devices driven by spin Hall torque. *Appl. Phys. Express* **7**, 053003 (2014).

19. Taniguchi, T. Different stochastic behaviors for magnetic field and current in domain wall creep motion. *Appl. Phys. Express* **7**, 053005 (2014).

20. Metaxas, P. J. *et al.* Creep and Flow Regimes of magnetic domain-wall motion in ultrathin Pt/Co/Pt films with perpendicular anisotropy. *Phys. Rev. Lett*. **99**, 217208 (2007).

21. Thiaville, A., Rohart, S., Jué, É., Cros, V. & Fert, A. Dynamics of Dzyaloshinskii domain walls in ultrathin magnetic films. *Europhys. Lett*. **100**, 57002 (2012).

22. Lemerle, S. *et al.* Domain wall creep in an ising ultrathin magnetic film. *Phys. Rev. Lett*. **80**, 849–852 (2007).

23. Kim, K.-J. *et al*. Interdimensional universality of dynamic interfaces. *Nature*. **458**, 740–742 (2009).

24. Slonczewskii, J. C. Theory of Bloch-line and Bloch-wall motion. *J. Appl. Phys.* **45**, 2705–2715 (1974).

25. Kim, S.-K., Lee, J.-Y., Choi, Y.-S., Guslienko, K. Y. & Lee, K.-S. Underlying mechanism of domain-wall motions in soft magnetic thin-film nanostripes beyond the velocity- breakdown regime. *Appl. Phys. Lett.* **93**, 052503 (2008).

26. Guslienko, K. Y., Lee. J.-Y. & Kim. S.-K. Dynamics of domain walls in soft magnetic nanostripes: topological soliton approach. *IEEE Trans. Mag*. **44**, 3079–3082 (2008).

27. Braun. H.-B. Topological effects in nanomagnetism: from superparamagnetism to chiral quantum solitons. *Adv. Phys.* **61**, 1–116 (2012).

28. Chetkin, M. V., Parygina, I. V., Roman, V. G. & Savchenko, L. L. Unidirectional collisions of vertical Bloch lines. *Sov. Phys. JETP* **78**, 93–97 (1994).

29. Wang, X. S., Yan, P., Shen, Y. H., Bauer, G. E. W. & Wang, X. R. Domain wall propagation thorough spin wave emission. *Phys. Rev. Lett*. **109**, 167209 (2012).

30. Wieser, R., Vedmedenko, E. Y. & Wiesendanger, R. *Phys. Rev. B* **81**, 024405 (2010).

31. Je, S.-G. *et al*. Asymmetric magnetic domain-wall motion by the Dzyaloshinskii-Moriya interaction. *Phys. Rev. B* **88**, 214401 (2013).

32. Nakatani, Y., Hayashi, N., Ono, T. & Miyajima, H. Computer simulation of domain wall motion in a magnetic strip line with submicron width. *IEEE Trans. Magn.* **37**, 2129–2131 (2001).

33. Fukushima, H., Nakatani, Y. & Hayashi, N. Volume average demagnetizing tensor of rectangular prisms. *IEEE Trans. Magn*. **34**, 193–198 (1998).

34. Hayashi, N., Saito, K. & Nakatani, Y. Calculation of demagnetizing field distribution

based on fast Fourier transform of convolution. *Jpn. J. Appl. Phys*. **35**, 6065–6073 (1996).

35. Rohart, S. & Thiaville, A. Skyrmion confinement in ultrathin film nanostructures in the presence of Dzyaloshinskii-Moriya interaction. *Phys. Rev. B* **88**, 184422 (2013).

36. Mizukami, S. *et al*. Gilbert Damping in Ni/Co Multilayer Films Exhibiting Large Perpendicular Anisotropy. *Appl. Phys. Express* **4**, 013005 (2011).

## Acknowledgements

This work was partly supported by a Grant-in-Aid for Scientific Research (S), a Grant-in-Aid for Scientific Research (C), a Grant-in-Aid for Young Scientists (B), a Grant-in-Aid for Scientific Research on Innovative Areas, Collaborative Research Program of the Institute for Chemical Research, Kyoto University, and R & D Project for ICT Key Technology of MEXT from the Japan Society for the Promotion of Science (JSPS).

## Author contributions

T.O. and K.-J.K. planned and supervised the study. Y.Y., K.-J.K., T.Taniguchi, T.Tono, K.U., and R.H. designed the experimental setup. Y.Y. fabricated the devices, performed the experiment, and collected data. K.Y. and Y.N. performed the simulation. Y.Y., K.-J.K, T.O., K.Y., and Y.N. analysed the data. Y.Y., K.-J K., T.M., and T.O wrote the manuscript. All authors discussed the results.

## Additional information

The authors declare no competing financial interests. Supplementary information accompanies this paper at www.nature.com/naturematerials. Reprint and permission information is available online at http://npg.nature.com/reprintsandpermissions.

Correspondence and requests for materials should be addressed to K.-J.K. or T.O.

**Figure Legends**

**Figure 1 | Schematic illustration of Walker breakdown (WB) phenomenon.** (**a**) Precessional domain wall (DW) motion in one-dimensional wire. (**b**) Formation of vertical Bloch line (VBL) in two-dimensional wire. (**c**) Four degenerate VBLs with different charges and chiralities.

**Figure 2 | Device structure with measurement setup and field-driven domain wall (DW) velocity.** (**a**) A scanning electron microscope image of the sample in a measurement configuration. Inset shows the schematic illustrations of magnetization states for the signal trace and the reference trace. (**b**) and (**c**) show the DW velocity as a function of the magnetic field $H$ for (**b**) symmetric Co/Ni and (**c**) asymmetric Co/Ni wires. Inset in (**b**) zooms into the DW velocity for the symmetric Co/Ni wire. Error bars are about the same size as the symbols (red dots). Solid lines in (**b**) and (**c**) are the calculated DW velocity with various values of $D$ based on the one-dimensional model. Experimentally determined material parameters are used. (**d**) and (**e**) show simulated DW velocity with various values of $D$ for (**d**) symmetric Co/Ni and (**e**) asymmetric Co/Ni wires. (**f**) Simulated DW velocity for an ideal wire. For all figures, the units of $D$ are mJ/m$^2$.

**Figure 3 | Snapshots of a moving domain wall (DW).** (**a**) and (**b**) show snapshots of a moving DW for (**a**) $D = 0$ mJ/m$^2$ and for (**b**) $D = 1.0$ mJ/m$^2$. The elapsed time between the panes in each figure is 0.2 ns for (**a**) and 0.1 ns for (**b**). The applied field is $|H| = 15$ mT for $D = 0$ mJ/m$^2$ (**a**) and $|H| = 150$ mT for $D = 1.0$ mJ/m$^2$ (**b**). The applied field pushes the DW to the right direction. The colour code for the in-plane magnetization component is shown by a colour wheel. The image height is 500 nm. (**c**) and (**d**) show schematic illustrations of the evolution of VBLs for the enclosed regions in (**a**) and (**b**).

**Figure 4 | Energy profiles and unidirectional collision of vertical Bloch lines (VBLs).** (**a**) Domain wall (DW) anisotropy energy, DMI energy, and total energy (from upper to lower panel) with respect to the azimuthal angle of the DW. (**b**) Schematic illustration of the evolution of a VBL under $D = 0$ mJ/m$^2$. The purple arrows indicate the local spin precession. (**c**) Schematic illustration of energy splitting of a degenerate VBL. (**d**) Schematic illustration of unidirectional collision of VBLs under a strong DMI. The purple arrows indicate the locking of the azimuthal angle $\varphi$ that results from the unidirectional collision of VBLs. (**e**) Time evolution of the average azimuthal angle of the DW above the WB. Strength of the DMI and magnetic field are denoted in the legend.

**Figure 5 | Quantification of the strength of the Dzyaloshinskii-Moriya interaction (DMI)** (**a**) Simulated domain wall (DW) velocity as a function of the in-plane longitudinal field $H_x$ for various values of $D$. (**b**) $H_{min}$ and $H_{DMI}$ as a function of $D$. (**c**) and (**d**) show experimental results of the DW velocity as a function of $H_x$ for (**c**) symmetric Co/Ni and (**d**) asymmetric Co/Ni wires, respectively. Blue and green symbols represent the up-down and down-up DW configuration as denoted in the inset of (**c**). Solid lines are the fitting curves obtained by parabolic fitting.

a

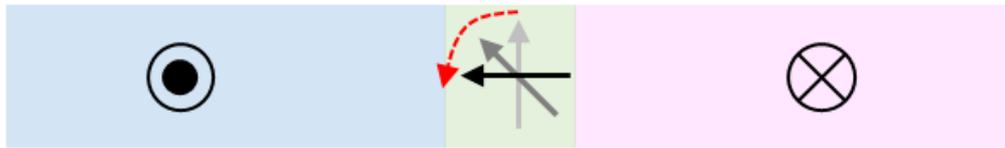

b

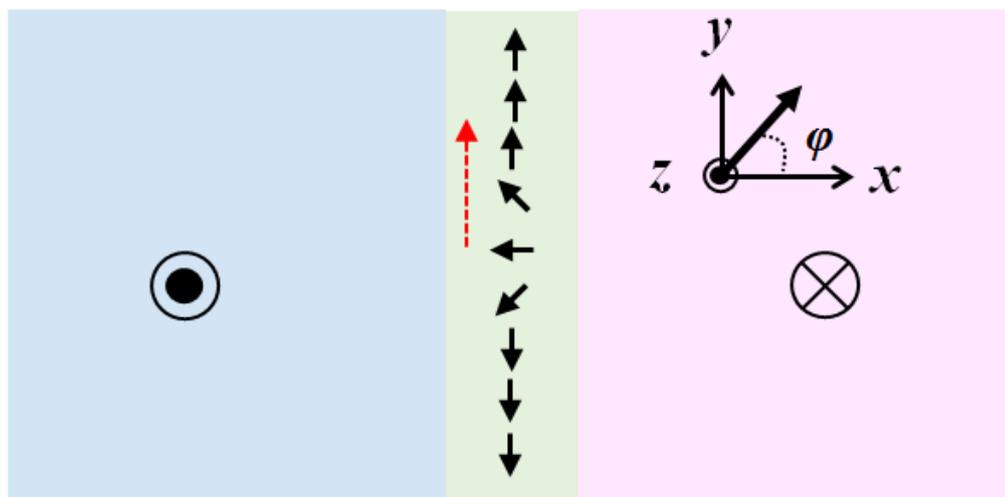

c

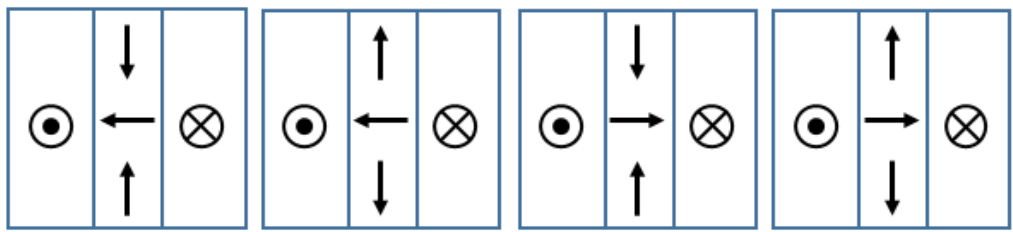

Fig. 1

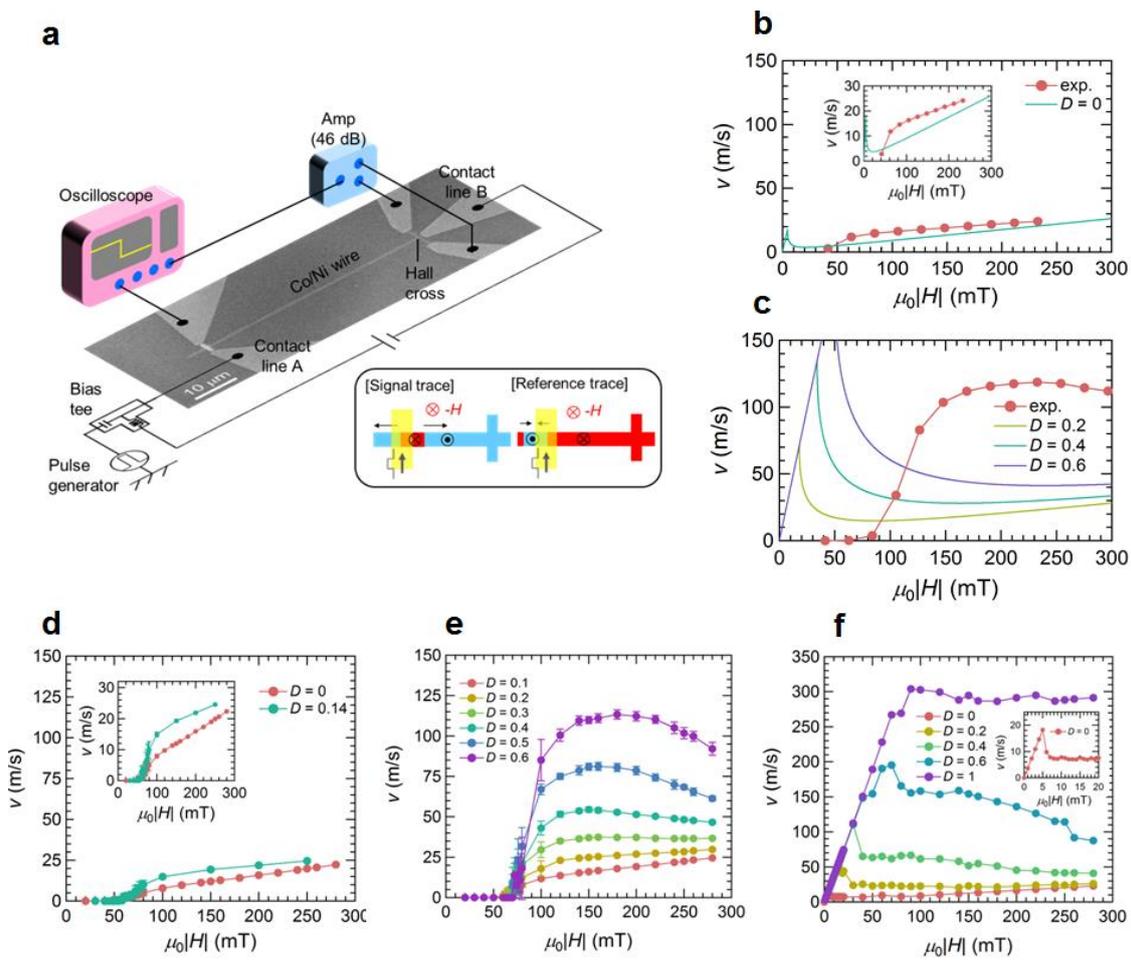

Fig. 2

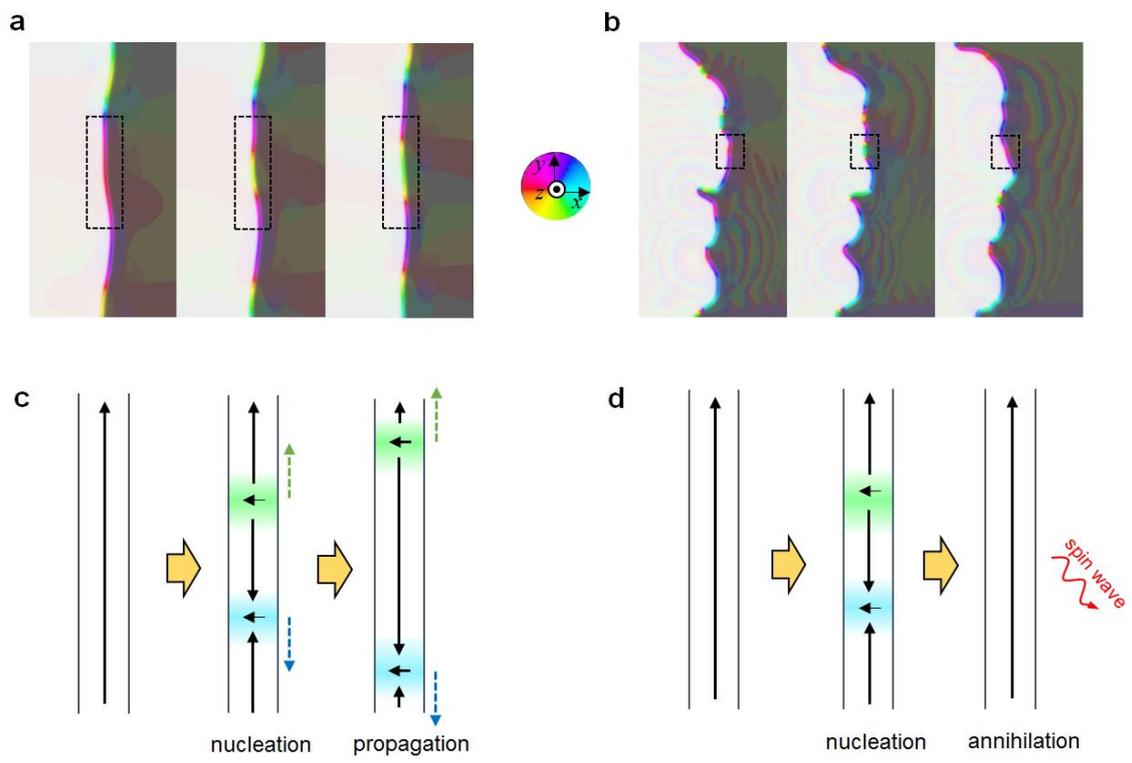

Fig. 3

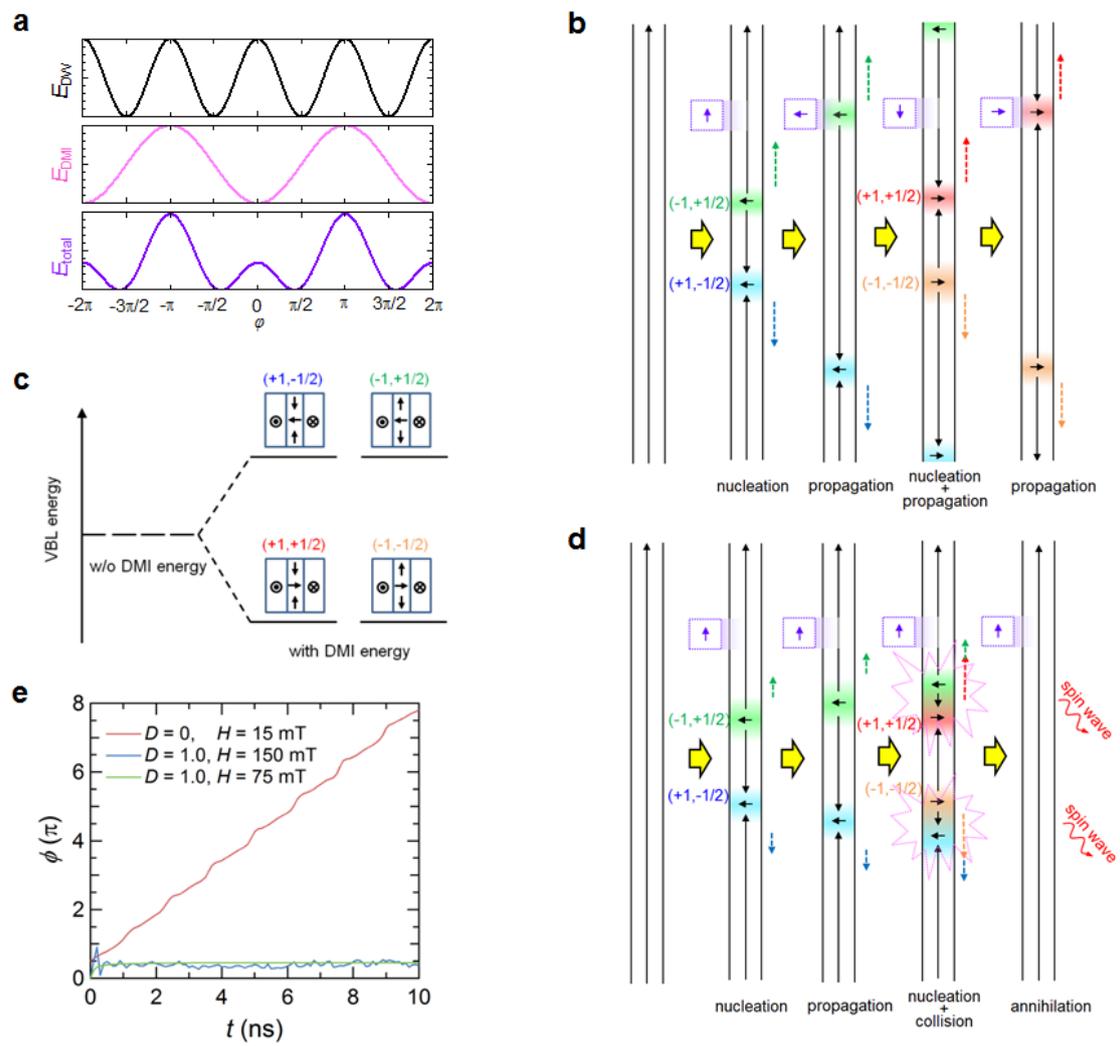

Fig. 4

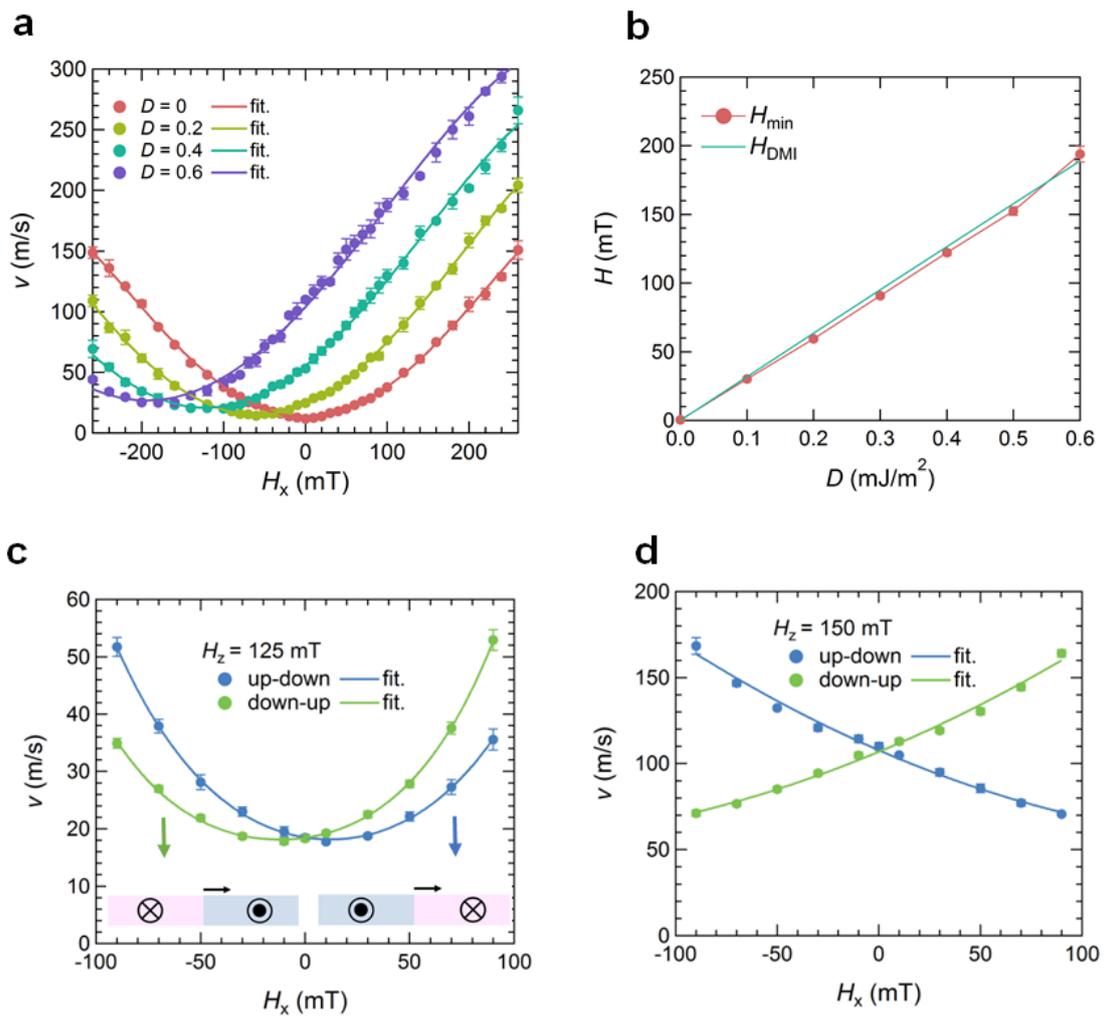

Fig. 5